\documentclass[%
  aps,          
  preprintnumbers,
  prl,          
  twocolumn,    
  showpacs,     
  superscriptaddress, 
  longbibliography,   
]{revtex4-2}

\usepackage{amsmath}        
\usepackage{amssymb}        
\usepackage{graphicx}       
\usepackage{dcolumn}        
\usepackage{bm}             
\usepackage{hyperref}       
\usepackage{xcolor}         
\usepackage{physics}
\usepackage{braket}
\usepackage{ulem}

\newcommand{\eqa}[1]{\begin{align}#1\end{align}}
\newcommand{\eq}[1]{\begin{equation}#1\end{equation}}

\newcommand{\eqsp}[1]{\begin{equation}\begin{split}#1\end{split}\end{equation}}

 \newcommand{\bea}{\begin{eqnarray}}
\newcommand{\eea}{\end{eqnarray}}
\newcommand{\be}{\begin{equation}}
\newcommand{\ee}{\end{equation}}
\newcommand{\ba}{\begin{align}}
\newcommand{\ea}{\end{align}}

\def\nli{\alpha}
\def\nlj{\beta}
\def\nlk{\gamma}

\def\nli{A}
\def\nlj{B}
\def\nlk{K}

\def\nla{A}
\def\nlb{B}
\def\nlm{M}
\def\cpi{i}
\def\cpj{j}
\def\nlkb{A}

\begin{document}
\preprint{CERN-TH-2026-208}

\title{A 1/4-BPS chiral algebra in D1D5:  from charge concentration to BPS chaos}

\author{Alexandre Belin}
\email{alexandre.belin@unimib.it}
\affiliation{Dipartimento di Fisica, Universit\`a di Milano - Bicocca, I-20126 Milano, Italy}
\affiliation{INFN, sezione di Milano-Bicocca, I-20126 Milano, Italy}

\author{Suzanne Bintanja}
\email{sbintanja@physics.ucla.edu}
\affiliation{Mani L. Bhaumik Institute for Theoretical Physics, Department of Physics and Astronomy, University of California Los Angeles, Los Angeles, CA 90095, USA}
\affiliation{CERN, Theory Division,
Geneva 23, CH-1211, Switzerland}

\author{Alessandra Gnecchi}
\email{alessandra.gnecchi@pd.infn.it}
\affiliation{INFN, Sezione di Padova, Via Marzolo, 8, 35131 Padova, Italy}

\author{Damian R. Musk}
\email{dmusk@caltech.edu}
\affiliation{Division of Physics, Mathematics and Astronomy, California Institute of Technology,  
1200 E. California Blvd., Pasadena, CA 91125, USA}

\author{Kyriakos Papadodimas}
\email{kyriakos.papadodimas@cern.ch}
\affiliation{CERN, Theory Division,
Geneva 23, CH-1211, Switzerland}

\begin{abstract}
The D1D5 CFT exhibits a charge concentration puzzle for BPS states, similar to, but simpler than, that of $\mathcal{N}=4$ SYM. We consider the chiral algebra of 1/4-BPS operators in the D1D5 CFT arising from a supercharge cohomology, and the resulting rational correlators.  Using crossing symmetry and this chiral algebra, we establish a non-perturbative proof of the existence of a family of 1/4-BPS states away from the charge concentration locus for general coupling and finite central charge. We further discuss possible implications for BPS chaos in the D1D5 CFT and compute a late-time BPS plateau at the orbifold point.
\end{abstract}


\maketitle

\section{Introduction}

The D1D5 CFT is one of the best-studied examples of holography, serving as a canonical example of AdS$_3$/CFT$_2$. Due to the high amount of supersymmetry, certain sectors of the theory are highly constrained, notably the $1/4$-BPS states, which provide a microscopic description of supersymmetric black holes \cite{Strominger:1996sh}. While the entropy of the microstates has been matched to great precision with the Bekenstein-Hawking formula (see \cite{Sen:2007qy} for a review)
the structure of these BPS states is not well understood. Outside of the signed count provided by the superconformal index, which is protected by supersymmetry and can be explicitly computed at the orbifold point \cite{Dijkgraaf:1996xw}, our knowledge of the structure of the 1/4-BPS sector is limited. This is particularly relevant because the 1/4-BPS sector encodes the microscopic degrees of freedom underlying both supersymmetric black holes and horizonless microstate geometries (see \cite{Bena:2022quy} for a review).

Recent progress has seen the emergence of refined characterizations of BPS states. First, the question of charge concentration has come into focus \cite{Kinney:2005ej, Cabo-Bizet:2018ehj, Choi:2018hmj, Benini:2018ywd, Larsen:2021wnu, Larsen:2024fmp, Chang:2024lxt, Choi:2025lck, Chang:2025wgo}.  In $\mathcal{N}=4$ SYM, this is the statement that BPS black holes in AdS$_5$ have to satisfy a non-linear constraint on their charges \cite{Kunduri:2006ek}, which poses a puzzle on whether BPS states exist away from this locus. In the D1D5 case, there is a similar charge constraint: BPS black holes must have right-moving $SU(2)$ $R$-charge $\widetilde{j}={k\over 2}$ in the NS sector, where $c=6k$ \cite{Breckenridge:1996is}. Second, BPS states have been separated into two types, monotones and fortuitous, encoding whether they correspond to a large backreacting gas of gravitons or to a true black hole \cite{Chang:2024zqi}. Microscopically, this is encoded in the $N$-dependence of the supersymmetry of the states. Finally, a new method was proposed for probing the chaotic properties of supersymmetric black holes \cite{Lin:2022rzw,Lin:2022zxd,Chen:2024oqv}, replacing standard probes such as energy level repulsion, which are not applicable to supersymmetric black holes with an exactly degenerate spectrum. 

A direct application of these refined characterizations of $1/4$-BPS states is difficult in the D1D5 CFT away from the orbifold point, due to our aforementioned limited knowledge on the structure of these states (see however 
\cite{Burrington:2012yq,Gaberdiel:2015uca,Hampton:2019oia,Guo:2019pzk,Guo:2020gxm,Guo:2020iua,Keller:2019suk,Benjamin:2021zkn,Apolo:2022fya,Guo:2022ifr,Hughes:2023rav, Hughes:2023uya,Gaberdiel:2024nge,liftingtoappear,Chang:2025rqy,Hughes:2025tdy,Chang:2025wgo,Hughes:2026naj,Giusto:2026rpl}). The goal of this paper is to make some progress on this front. We consider the chiral algebra of 1/4-BPS states corresponding to cohomology classes of a supercharge. This is similar to chiral algebras emerging in four-dimensional SCFTs \cite{Beem:2013sza}. Correlators within this algebra are drastically simplified; they are meromorphic, rational functions. Studying crossing symmetry of these correlators, we establish the existence of a class of 1/4-BPS states \textit{away} from the charge concentration locus, for all values of the coupling and finite central charge $c$. We also discuss some implications for BPS chaos in the D1D5 CFT.

\section{The chiral algebra of 1/4 BPS states}

Consider one of the right-moving supercharges, $\widetilde{G}_{-{1\over 2}}^{++}$ of a  ${\cal N}=(4,4)$ SCFT, in the NS sector. It obeys 
\be
(\widetilde{G}_{-{1\over 2}}^{++})^2=0\,,
\ee
and we can consider its cohomology. It is generally infinite-dimensional, and its elements correspond to  1/4-BPS operators with $SU(2)_R$ quantum numbers aligned as $\widetilde{j}^3=\widetilde{j}$. The OPE of two such 1/4-BPS operators has the form
\begin{align}
\label{14bpsope}
   \Phi_{\nli}(z,\overline{z}) \!\cdot \!\Phi_{\nlj}(0,0)\! = \!\!\sum_K C_{\nli \nlj}^{\nlk} & {\Phi_{\nlk}(0,0) \over z^{h_{\nli}+h_{\nlj}    -h_{\nlk}} }  \cr & + [\widetilde{G}_{-{1\over 2}}^{++}, {\rm other}\}(z,\overline{z})\,.
\end{align}
The states $\Phi_\nlk$ on the RHS are also 1/4-BPS, while all other operators are $\widetilde{G}_{-{1\over 2}}^{++}$ exact. The first term has no $\overline{z}$-dependence, since $SU(2)_R$ charge conservation and the BPS bound imply $\widetilde{h}_{\nlk}=\widetilde{h}_{\nli}+\widetilde{h}_{\nlj}$. Hence, the OPE defines a chiral algebra of multiplication of cohomology classes of $\widetilde{G}_{-{1\over 2}}^{++}$. The OPE coefficients $C_{\nli \nlj}^{\nlk}$ 
encode the structure of this algebra 
and provide a generalization of the 3-point functions of the usual chiral ring of $1/2$-BPS operators \cite{Lerche:1989uy}. In \cite{deBoer:2008ss, Baggio:2012rr} it was proven that in ${\cal N}=(4,4)$ SCFTs both 1/2- and  1/4-BPS OPE coefficients $C_{\nli \nlj}^{\nlk}$ are covariantly constant on the conformal manifold \footnote{The theorem applies to 1/4-BPS states that do not lift under marginal deformations.}.

The algebra \eqref{14bpsope} generated by the 1/4-BPS states has the structure of a vertex operator algebra. 
 It is graded by left-conformal dimension and the left and right $R$-charge. We denote by ${\cal A}^{\widetilde{j}}$ the elements of a given right $R$-charge.
As a vector space, the algebra is   
$
{\cal A}={\cal A}^0 \oplus {\cal A}^{1\over 2}\oplus {\cal A}^1 ... \oplus {\cal A}^{c\over6}\,.
$
Under OPE multiplication
$
\label{su2grading}
{\cal A}^{\widetilde{j}_1} \otimes {\cal A}^{\widetilde{j}_2} \rightarrow {\cal A}^{\widetilde{j}_1+\widetilde{j}_2}\,.
$
Finally, via spectral flow, we have 
$
{\cal A}^{c\over 6} \cong{\cal A}^0
$. The sector ${\cal A}^0$  corresponds to 
the set of operators generated by the left $
\mathcal{N}=4$ SCA generators \footnote{Here we assume that there are no accidental conserved currents.}. The higher degree subspaces ${\cal A}^j$ contain non-trivial 1/4 BPS states.

An immediate consequence of the OPE structure \eqref{14bpsope} is that certain correlators of 1/4-BPS operators are {\it meromorphic, rational} functions of the coordinates. 
To see this, consider a correlation function of $n$ $1/4$-BPS operators, all aligned to satisfy $\widetilde{j}_{{\nli}_i}^3=\widetilde{j}_{\nli_i}$, with one ``conjugate" 1/4-BPS state with $\widetilde{j}_\nlj^3=-\sum_i \widetilde{j}_{\nli_i}$
\be
\label{cor14bps}
F(z_i,\overline{z}_i)\equiv\langle \Phi^\dagger_\nlj(\infty) \,\,\Phi_{\nli_1}(z_1,\overline{z}_1)...\Phi_{\nli_n}(z_n,\overline{z}_n) \rangle\,.
\ee
We have
\be
    {\partial F \over \,\,\partial\overline{z}_{\nli_i}} \!=\!\langle \Phi^\dagger_\nlj(\infty)\!\cdots\![\widetilde{L}_{-1},\Phi_{\nli_i}](z_i,\overline{z}_i)\!\cdots\!\Phi_{\nli_n}(z_n,\overline{z}_n)\rangle.
\ee
The superconformal algebra implies $
\widetilde{L}_{-1}= \{\widetilde{G}_{-{1\over 2}}^{++},\widetilde{G}_{-{1\over 2}}^{--}\}.
$
The state $\Phi_{\nli_i}$ is 1/4-BPS, and hence annihilated by $\widetilde{G}^{++}_{-{1\over 2}}$, so by using 
a standard superconformal Ward identity \cite{deBoer:2008ss},  which allows us to move the supercharge $\widetilde{G}^{++}_{-{1\over 2}}$ away from $z_i$, where it annihilates all other operators and gets no contribution from infinity, we find 
\be
{\partial F \over \,\,\partial\overline{z}_{\nli_i}}=0\,.
\ee
Repeating for all arguments, using the OPE expansion at coincident points and behaviour at infinity, we conclude that $F$ is a rational function. Moreover, the aforementioned non-renormalization theorem implies that $F$ does not depend on the couplings \footnote{This is expected to hold for 1/4-BPS insertions that do not lift. In combination with continuity of correlators under marginal deformations, it follows that at the orbifold point, intermediate accidental 1/4-BPS states must give a vanishing contribution to these correlators.}.

Consider now a 4-point function of two 1/2-BPS operators $\phi_\cpi,\phi_\cpj$, and two 1/4-BPS operators $\Phi_\nla,\Phi_\nlb$. From the previous discussion, we have
\be
\label{4point12bps}
F(z) \equiv \langle \Phi^\dagger_\nlb(\infty) \,\,\phi_{\cpi}(z,\overline{z})\,\phi_\cpj(1)
\,
\,\Phi_\nla(0)\rangle ={P(z) \over z^r}\,,
\ee
where $P(z)$ is a polynomial and $r\in {\mathbb Z}_{\geq 0}$ \footnote{There is no singularity at $z=1$ because the OPE of the two 1/2-BPS operators is regular.}.  

We will now consider the conformal block expansion in the three channels $z=0,1,\infty$. 
First, we consider the channel $z\rightarrow 0$. We have the expansion
\be
\label{channel0}
F(z) = \sum_{M} C_{\cpi \nla}^\nlm\, C_{\cpj \nlm}^\nlb \, {1\over z^{h_\cpi +h_\nla}} \,{\cal G}_{0}^{h_\nlm}(z)\,,
\ee
where the conformal block in this channel is   
${\cal G}_0^{h_\nlm}(z)=z^{h_\nlm}\,_2F_1(h_\nlm - h_\nla+h_\cpi,h_\nlm+h_\cpj-h_\nlb,2h_\nlm,z)\,.
$
Second, in the $z\rightarrow \infty$ channel we have
\be
\label{channelinfty}
F(z)=\sum_{M} \, C_{\cpj \nla}^\nlm\, C_{\cpi \nlm}^\nlb\,\,z^{h_\nlb-h_\cpi} {\cal G}_{\infty}^{h_\nlm}(1/z)\,,
\ee
where 
$
{\cal G}_{\infty}^{h_\nlm}(z) = z^{h_\nlm}\,_2F_1(h_\nlm - h_\nla+h_\cpj,h_\nlm + h_\cpi-h_\nlb,2h_\nlm,z)\,.
$
Finally, in the  $z\rightarrow 1$ channel, we have
\be
\label{channel1}
F(z) = \sum_{M} C_{\cpi \cpj}^\nlm C_{\nla \nlm}^\nlb \,{1\over (z-1)^{h_\cpi+h_\cpj}}\,\,{\cal G}_{1}^{h_\nlm}(1-z)\,,
\ee
where the conformal block takes the form
$
{\cal G}_{1}^{h_\nlm}(z) = z^{h_\nlm} \,_2F_1(h_\nlm-h_\cpj+h_\cpi,h_\nlm+h_\nla-h_\nlb,2h_\nlm,z)\,.
$

Crossing symmetry implies that \eqref{channel0}, \eqref{channelinfty}, and \eqref{channel1} must be equal and must all be compatible with \eqref{4point12bps}.
One can see that the only way that this can be satisfied is if in each channel: i) We have a finite number of primaries contributing ii) Their conformal dimensions are such that the conformal blocks truncate to polynomials. More systematically, we obtain the following implications:

\begin{enumerate}
 \item If $h_\nlb = h_\nla+h_\cpi + h_\cpj +n$, $n\in {\mathbb Z}_{\geq 0}$, then
 $
 F(z) = P_n(z)
 $
 where $P_n$ is a polynomial of degree $n$. 
\item If $h_\nlb = h_\nla -h_\cpi -h_\cpj -n$, $n\in {\mathbb Z}_{\geq 0}$, then 
$
 F(z) = {1\over z^{2h_\cpi+n}}P_n(z)
$
 where $P_n$ is a polynomial of degree $n$. 
 \item If $h_\nla-h_\cpi -h_\cpj< h_\nlb < h_\nla + h_\cpi + h_\cpj$, then
 $
 F(z)=0
 $
 since it is impossible to satisfy crossing symmetry.
 \end{enumerate}

We note some additional consequences of crossing symmetry. First, we get predictions for the vanishing of certain sums of 1/4-BPS OPE coefficients, for instance in the first case above 
\be
\label{sela}
\sum_{\overline{M}} C_{\cpi \nla}^\nlm C_{\cpj \nlm}^\nlb =0 \qquad {\rm if}\qquad  h_\cpj+h_\nlm > h_\nlb\,.
\ee
where the notation $\overline{M}$ means that the sum is now over operators of fixed dimension $h_M$.
Second, we get non-trivial relations between generally non-vanishing OPE coefficients of the chiral algebra. For example, in the case 1. above, and in the simplest case with $n=0$, we find
\be
\label{crosssymc}
\sum_{\overline{M}} C_{\cpi \nla}^\nlm C_{\cpj \nlm}^\nlb = \sum_{\overline{N}} C_{\cpj \nla}^N C_{iN}^B = \sum_{\overline{K}} C_{ij}^K C_{KA}^B\,,
\ee
where $h_\nlm = h_\cpi+h_\nla, h_N = h_\cpj+h_\nla , h_K=h_\cpi+h_\cpj$ (and in this case $M,N,$ and $K$ are also chiral primaries). It would be interesting to analyze these constraints more systematically and also study crossing symmetry of four 1/4-BPS operators.

\section{A class of non-concentrated 1/4 BPS states in \texorpdfstring{$\boldsymbol{{\cal N}=(4,4)}$}{N44} theories}
\label{sec:proofnoncc}

The analysis of the previous section holds even in a $\mathcal{N}=(2,2)$ SCFT, though with just $(2,2)$ SUSY it is unclear whether most ${\cal A}^j$ are non-empty at finite coupling. 
We now use the extended ${\cal N}=(4,4)$ structure of the D1D5 CFT to demonstrate that  
its chiral algebra is indeed non-trivial. 
In particular, using crossing symmetry, we prove the existence of families of 1/4-BPS states populating the entire spectrum of $0\leq \widetilde{j} \leq k$, going beyond those predicted by the elliptic genus and representation theory \footnote{We emphasize that at this point we are not making any statement about the entropy and monotone/fortuitous nature of these states.}. The following analysis holds for all values of the coupling and any $c=6k$.

Let us consider an operator $\Phi_\nlkb$ which is either 1/2- or 1/4-BPS and $\phi_\cpi$ to be a 1/2-BPS operator of dimension $(1/2,1/2)$, with $R$-charge alignment $(1/2,1/2)$. Consider also the element in the same 1/2-BPS multiplet defined as 
$
\chi_\cpi=J_0^- \phi_\cpi\,,
$
which also has dimension $(1/2,1/2)$ and charges $(-1/2,1/2)$. Then we consider the combination
\begin{align}
\label{deffung}
G(z) \equiv& \,\langle \Phi_\nlkb^\dagger(\infty)\,\, \phi_\cpi(z,\overline{z}) \,\,\phi_\cpi^\dagger(1)\,\,\Phi_\nlkb(0)\rangle\cr & -\overline{z}\,\langle \Phi_\nlkb^\dagger(\infty) \,\,\chi_\cpi^\dagger(z,\overline{z}) \,\,\chi_\cpi(1)\,\,\Phi_\nlkb(0)\rangle\,.
\end{align}
Using ${\cal N}=(4,4)$ superconformal Ward identities reviewed in the supplemental material we find
\be
\label{deffung2}
{\partial G(z) \over \partial {{\overline z}}}=0 \,,
\ee
so $G(z)$ is a meromorphic, rational function. 

Expanding in conformal blocks in the $z=0$ channel, and using identity \eqref{cblockidentity}, we find that only intermediate 1/4-BPS states can possibly contribute, as 
\be
\label{g14bps}
\hspace{-27pt}G(z)\!=\! {1\over z^{h_\nlkb+{1\over 2}}}\! \sum_{{\cal O}_\nlm}\!\left(\,C_{\phi_i \Phi_\nlkb}^{{\cal O}_\nlm}  C_{{\cal O}_\nlm \phi_i^\dagger}^{\Phi_\nlkb^\dagger}\!  - \!\,C_{\chi_i^{\dagger} \Phi_\nlkb}^{{\cal O}_\nlm} C_{{\cal O}_\nlm \chi_i}^{\Phi_\nlkb^\dagger}\right)\!{\cal G}^{h}(z),
\ee
where the first/second sum is over 1/4 BPS states ${\cal O}_\nlm$ of right-moving charge $\widetilde{h}= \widetilde{h}_\nlkb+{1\over 2}$ and $\widetilde{h} = \widetilde{h}_\nlkb-{1\over 2}$ respectively. However, a priori it is unclear if 1/4-BPS states with these quantum numbers exist; a logical possibility is that all OPE coefficients vanish and $G(z)=0$.

To exclude this possibility, we consider the expansion in the $z=1$ channel. The relevant terms come from the exchange of the identity and $\widetilde{J}^3$ and we find
\be
\label{geng1}
G(z) =\left(1-{2\widetilde{j}_\nlkb \over k}\right){1\over 1-z} +{P(z) \over z^r}\,,
\ee
where we used $c=6k$. 
Examining crossing symmetry from $z=0$ and $z=\infty$
, we encounter a situation similar to case iii) of the previous section, which implies that
\be
\label{polg}
P(z)=0\,.
\ee

If the state $\Phi_\nlkb$ is away from the charge concentration locus, say with $\widetilde{j}_\nlkb<{k\over 2}$, then combining \eqref{geng1} and \eqref{polg} we can predict that $G(z)\neq 0$. Then, using \eqref{g14bps}, there is a family of non-concentrated 1/4-BPS states with $(j,\widetilde{j})=(j_\nlkb+{1\over 2},\widetilde{j}_\nlkb+{1\over 2})$ and left-moving dimension $h_\nlkb = j_\nlkb + {1\over 2} +n$, $n={\mathbb Z}^+$. Moreover, we find  
the OPE coefficients 
\be
\begin{split}
&\sum_{{\cal O}_\nlm}|C_{\phi \Phi_\nlkb}^{{\cal O}_\nlm}|^2-|C_{\chi_i^{\dagger} \Phi_\nlkb}^{{\cal O}_\nlm}|^2\\
&\hspace{40pt}= \left(1-{2 \widetilde{j}_\nlkb \over k}\right) {\Gamma(2h_\nlkb + n)^2\over \Gamma(2h_\nlkb)\Gamma(2h_\nlkb+2n)}\,,
\end{split}
\ee
where $n$ is controlled by  $h_\nlkb= j_\nlkb + {1\over 2} +n$.  

Notice that this analysis can be applied recursively to the newly established 1/4-BPS states, as well as by using other $(1/2,1/2)$ primaries $\phi_i$, to generate an interesting span of 1/4-BPS states with general $\widetilde{j}$, which intuitively correspond to the ``seed" state $\Phi_A$, decorated by a gas of chiral primaries. One possible starting ``seed" state $\Phi_A$ for this construction is
a non-charge concentrated 1/2-BPS state, whose existence is guaranteed for all values of the coupling. Using the prescription above, we then generate a whole family of 1/4-BPS states away from the charge-concentration locus at all coupling, whose existence does not follow from the elliptic genus.

On the other hand, we cannot predict the entropy or properties of these states. For example, it is a logical possibility that they are all monotone and correspond  
to fuzzball geometries in AdS$_3$ \cite{Chang:2024zqi, Chang:2025rqy, Hughes:2025tdy}. 
Our construction provides 1/4-BPS states through protected OPEs with chiral primaries. 
Some of these states may overlap with superstrata \cite{Bena:2015bea,Shigemori:2020yuo}  or more general microstate geometries, such as bubbling solutions, whose precise CFT duals are not known \cite{Bena:2025pcy}.

\section{Comments on BPS Chaos}

We now present some observations related to BPS chaos. Consider the LMRS observable
\be
\label{projop}
\widetilde{\cal O}= P_{\textrm{BPS}_1} {\cal O} P_{\textrm{BPS}_2} \,,
\ee
where $P_{\textrm{BPS}_{1,2}}$ are projectors onto 1/4 BPS subspaces and ${\cal O}$ is a choice of simple operator which we will shortly specify. The nature of the eigenspectrum of this operator was proposed as a diagnostic for chaos in the BPS subspace \cite{Lin:2022zxd,Lin:2022rzw,Chen:2024oqv}. However, in the D1D5 CFT, there are subtleties in selecting non-trivial probes.  

At strong coupling, single-trace operators of low dimension are in short multiplets. $R$-symmetry neutral operators are super-descendants, and their 3-point functions with BPS states vanish due to superconformal Ward identities.  We can have non-trivial 3-point functions for single-trace chiral primary probes, but since those are $R$-charged, the projectors $P_{\textrm{BPS}_{1,2}}$ must have \textit{different} $R$-charges. If there only existed 1/4 BPS states on the charge concentration locus, $\widetilde{\cal{O}}$ would vanish. 
Thankfully, given the results of the previous section,
the LMRS operator $\widetilde{\cal O}$ is non-trivial.

A second issue is the non-renormalization theorem of three-point functions \cite{Baggio:2012rr}. 
Because of the high number of additional selection rules at the orbifold point, this theorem would suggest there cannot 
be strong chaos for these probes, even at strong coupling. However, it is important to take into account the lifting of operators.
In the D1D5 CFT (as opposed to $\mathcal{N}=4$ SYM), the leading asymptotic growth of the entropy of BPS states is the same at weak and strong coupling, given by Cardy's formula \cite{cardyformula}. Including prefactors, we expect (at least for charge-concentrated states) the following number of BPS states (in the Ramond sector)
\be
\rho_{\text{BPS}}(h) = \alpha(h/c)\left(\frac{c}{(h-\frac{c}{24})^3}\right)^{\frac{1}{4}} e^{2\pi \sqrt{\frac{c}{3}(h-c/24)}} \,.
\ee
where $\alpha(h/c)$ scales like $c^0$ when $c\gg 1$ and $h\gtrsim c$. It encodes the number of BPS light states in the bulk. For the elliptic genus, one can show that $\alpha_{\textrm{SUGRA}} < \alpha_{\textrm{orbifold}}$ \cite{Apolo:2024kxn}. By analogy, we expect an order one fraction of the BPS states to lift as we deform away from the symmetric orbifold.  Therefore, even if absent at the orbifold point, chaos can emerge by restricting \eqref{projop} to the small subset of 1/4-BPS states that do not lift \footnote{It is sufficient that an order one fraction of the states lifts for there to be strong chaos for the LMRS operator \citep{Belin:2026ssc}.}. Another possibility is that the non-concentrated states of the previous subsection correspond to BPS hairy configurations. In that case, the natural expectation is that the chiral primary probe corresponds to a hairy excitation and does not probe the near-horizon JT gravity region. This would imply no BPS chaos for chiral primary probes. Distinguishing between these possibilities goes beyond the scope of this work.

It is interesting to consider the second moment $\Tr \widetilde{\cal O} \widetilde{\cal O}^{\dagger}$ of the projected operator, related to the late-time limit of a Lorentzian 2-point function on a BPS black hole \cite{Lin:2022rzw,Lin:2022zxd}. This quantity can be obtained from the Euclidean torus two-point function
\eqsp{
&\left\langle {\cal O}(z,\bar{z}) {\cal O}^\dagger(0,0)\right\rangle_\tau\\
&\hspace{25pt}= \Tr_R\left[q^{L_0-c/24}\bar{q}^{\bar{L}_0-c/24} {\cal O}(z,\bar{z}) {\cal O}^\dagger(0,0)\right]\,,
}
where, as usual, $q= e^{2\pi i\tau}$, and $z=\phi+it_E$ \footnote{Note that this two-point function is not normalized for the moment.}. Moreover, the modular parameter $\tau$ encodes the left- and right-moving temperature as $\beta_L=-2\pi i\tau$ and $\beta_R=2\pi i\bar{\tau}$. To project to the BPS subspace, we take the extremal limit $\beta_R\rightarrow\infty$. After inserting a complete set of states and continuing to Lorentzian time via $t=it_E$, we can perform a late-time average to extract the late-time plateau \footnote{Strictly speaking, the time integral diverges due to lightcone singularities. The divergence can be regulated by deforming the integration contour into the complex plane 
and using a minimal subtraction regularization scheme. In practice, this means that in the sum over OPE coefficients we only keep the $t$-independent terms. Another possible regularization would be to consider a two-sided correlator in the extremal limit.}\footnote{The appropriate normalization factor that should be included is the partition function in the extremal limit $ Z(c,\beta_L)= \sum_{\psi \in \mathcal{H}_{\frac{1}{4}\text{-BPS}}} e^{-\beta_L (h-\frac{c}{24})}$.}
\eqa{ \label{finaleqlatetime2dCFT}
\!\!F(c,{\cal O},\beta_L)&\equiv\lim_{T\rightarrow\infty}\frac{1}{T}\int_0^T dt\int_0^{2\pi}d\phi \left\langle {\cal O}(t,\phi) {\cal O}^\dagger(0,0)\right\rangle_{\beta_R\rightarrow\infty}\notag\\
&\hspace{-12pt}=
\hspace{-10pt}\sum_{ \substack{\psi,\psi'\in\mathcal{H}_{\frac{1}{4}\text{BPS}}\\h=h'}}\hspace{-13pt}e^{-\beta_{L}(h-c/24)}\braket{\psi | {\cal O} | \psi'} \braket{\psi'\vert {\cal O}^{\dagger}\vert \psi}\,.
}
We perform this calculation at the orbifold point (and since the calculation below is valid in any symmetric orbifold, we keep the central charge of the seed generic). 
Using the methods of \cite{Belin:2025nqd}, we can write the torus two-point function at the orbifold point in terms of seed quantities. Taking the probe operator to be untwisted and single-trace, 
the orbifold plateau takes the following form
\begin{widetext}
\eq{
F_N(c,{\cal O}_{\text{seed}},\beta_L)=\frac{1}{N}\sum_{\substack{g_{1,2}\in S_N\\g_1g_2=g_2g_1}}\sum_{a,b,d}\hspace{0pt}\delta_{a,1}d^{2(1-\Delta_{{\cal O}})}\hspace{-15pt}\sum_{\substack{\psi,\psi'\in\frac{1}{4}\text{-BPS}\\ h=h'}} \hspace{-15pt} e^{-\frac{\beta_L+b}{d}(h-c_{\text{seed}}/24)}\abs{\bra{\psi}{\cal O}_{\text{seed}}\ket{\psi'}}^2\hspace{-20pt}\prod_{\substack{(a',b',d')\neq(a,b,d)}}\hspace{-20pt}Z_\text{seed}\left(\frac{a'\beta_L+b'}{d'}\right)\,.\label{eq:orbplat}
}
\end{widetext}
Here, the sum over commuting group elements is the sum over all possible covering surfaces of the torus. This is a product of tori with modular parameter $\frac{a\beta_L+b}{d}$, and the parameters $(a,b,d)$ are determined by the orbits of $g_{1,2}$ in the abelian group generated by $g_1$ and $g_2$. The delta function $\delta_{a,1}$ is present to ensure that the seed operators can only be inserted on primitive coverings with the appropriate monodromy structure. The details of the calculation are presented in the supplemental material. Assuming that long twisted cycles dominate, as for the partition function and thermal 2-point function \cite{Keller:2011xi,Belin:2025nqd}, we can write the (normalized) orbifold plateau in terms of the seed plateau as
\eq{
\label{eq:14plateauhyp}
\frac{F_N(c_{\text{seed}},{\cal O}_{\text{seed}},\beta_L)}{Z_N(c_{\text{seed}},\beta_L)}\sim\frac{F(c_{\text{seed}},{\cal O}_{\text{seed}},\beta_L/N)}{N^{2\Delta_{{\cal O}}}e^{\pi^2 c_{\text{seed}} N/3\beta_L}}\,.
}
A more rigorous derivation is possible when taking also $\beta_L\rightarrow\infty$, which restricts to the intermediate 1/2-BPS subspace. In this case, several twisted sectors contribute, and the answer is fixed by $S_N$ combinatorics, which leads to a $\sqrt{N}$ suppression compared to the seed plateau
\eq{
\hspace{-12pt}\frac{F_N(c,{\cal O}_{\text{seed}},\beta_L=\infty)}{Z_N(c,\beta_L=\infty)}\!=\! \frac{\gamma(\Delta_\mathcal{O})}{\sqrt{N}}\frac{F(c,\mathcal{O}_{\text{seed}},\beta_L=\infty)}{Z(c,\beta_L=\infty)}.
}
Here $\gamma(\Delta_\mathcal{O})$ is an order one coefficient for $\Delta_{\mathcal{O}}>1$ \footnote{When $\Delta_{\mathcal{O}}=\frac{1}{2}$, $\gamma$ is enhanced and scales linearly with $N$, while for $\Delta_{\mathcal{O}}=1$ $\gamma$ gives a $\log(N)$ scaling.}.
Since 1/2-BPS states do not lift, and using the non-renormalization theorem, this is a prediction for the 1/2-BPS plateau that is valid even at strong coupling. It would be interesting to reproduce this result from the bulk by considering the late-time 2-point function on the appropriate ensemble of 1/2-BPS fuzzball geometries, along the lines of \cite{Balasubramanian:2005qu,Balasubramanian:2016ids,Bombini:2018aop,Bena:2019azk}. 
\section{Discussion}
In this paper, we have introduced a chiral algebra of 1/4 BPS states for two-dimensional CFTs with $\mathcal{N}=(4,4)$ supersymmetry. We exploited this algebra, together with crossing symmetry of four-point functions, to establish the existence of CFT operators whose quantum numbers do not lie on the charge concentration locus ($\widetilde{j}={k\over 2}$). We discussed the relevance of this fact for the study of BPS chaos in D1D5 CFT. 

The most pressing open question is the bulk understanding of these non-concentrated states. In AdS$_5$, they are conjectured to be given by grey galaxies: bulk geometries consisting of a core black hole dressed with additional matter \cite{Kim:2023sig,Bajaj:2024utv}. Such geometries have been conjectured to exist also in AdS$_3$ \cite{Larsen:2025jqo}. However, to the best of our knowledge, they have not been explicitly constructed (see, however, the interesting geometries of \cite{Bena:2011zw}). We hope to return to this question in the future.

\section*{Acknowledgments}
We would like to thank Alejandra Castro, Sean Colin-Ellerin, Seok Kim, Shota Komatsu, Finn Larsen, Wolfgang Lerche, Shiraz Minwalla, Sameer Murthy, Palash Singh, and Alberto Zaffaroni for discussions. We would like to especially thank Monica Guica for collaboration during the initial stages of this project. The work of SB is supported by the Mani L. Bhaumik Institute for Theoretical Physics.

\bibliographystyle{apsrev4-2}   
\bibliography{ref}

\onecolumngrid  
\setcounter{equation}{0}
\setcounter{figure}{0}
\setcounter{table}{0}
\renewcommand{\theequation}{S\arabic{equation}}
\renewcommand{\thefigure}{S\arabic{figure}}
\renewcommand{\thetable}{S\arabic{table}}

\begin{center}
  \textbf{\large Supplemental Material}\\[4pt]
\end{center}

\section{Proof of meromorphicity of $G(z)$}

We start with a 1/4- or 1/2-BPS state $\Phi_K$ and a 1/2-BPS state $\phi_i$ of dimension $(1/2,1/2)$. We also define $\chi_i =J_0^-\phi_i$. Then consider the 4-point function
\be
I=\langle \Phi_K^\dagger(\infty) \,\,\widetilde{G}^{+A}_{-{1\over 2}} \chi_i^\dagger (z) \,\, \widetilde{G}_{-{1\over 2}}^{+B} \phi_i^\dagger(y)\,\,\Phi_K(0)\rangle\,,
\ee
where $A,B = \pm$ correspond to the $SU(2)_{\rm outer}$ sector of the ${\cal N}=4$ SCA.
The operator $\Phi_K$ is annihilated by $\widetilde{G}^{+A}$. Hence we can move the supercharges from the operator at $z$ to $y$ or the other way around. We do not get a contribution from the operator at infinity.
We will also need the following identity following from the ${\cal N}=4$ SCA, for chiral primaries $\phi_i$ of weight $(1/2,1/2)$
\be
\label{anotherid}
\{\widetilde{G}^{+A}_{-{1\over 2}},[\widetilde{G}^{+B}_{-{1\over 2}} ,\phi_i^\dagger]\} =\epsilon^{AB} \partial_{\overline{z}} \, [\widetilde{J}_0^+,\phi_i^\dagger]\,,
\ee
and an identical one for the field $\chi_i^\dagger$. Moving the supercharge away from $y$ we get 
\be
I = \epsilon^{AB} \partial_{\overline{z}}\,\langle \Phi_K^\dagger(\infty) \widetilde{J}_0^+ \chi_i^\dagger(z) \,\,\phi_i^\dagger(y)\,\,\Phi_K(0)\rangle\,,
\ee
where we used the ${\cal N}=4$ SCA. Now using that $\chi_i =J_0^-\phi_i$ 
we find
\be
\label{expr1}
I= -\epsilon^{AB} \partial_{\overline{z}}\,\langle \Phi_K^\dagger(\infty) \phi_i(z) \,\,\phi_i^\dagger(y)\,\,\Phi_K(0)\rangle\,.
\ee
Similarly, we can move the supercharge away from $z$ to get
\be
\label{expr2}
I = -\epsilon^{BA} \partial_{\overline{y}} \langle \Phi_K^\dagger(\infty) \, \chi_i^\dagger(z)\,\, \chi_i(y) \,\Phi_K(0)\rangle\,.
\ee
Now we use a conformal Ward identity
\be
\label{confward}\partial_{\overline{y}} \langle \Phi_K^\dagger(\infty) \, \chi_i^\dagger(z)\,\, \chi_i(y) \,\Phi_K(0)\rangle \\
= - \partial_{\overline{z}} \left({\overline{z}\over \overline{y}} \langle \Phi_K^\dagger(\infty) \, \chi_i^\dagger(z)\,\, \chi_i(y) \,\Phi_K(0)\rangle \right)\,,
\ee
which holds for any 2d CFT  provided that the internal operators $\chi$ have dimension $(1/2,1/2)$. 
Using the two expressions \eqref{expr1}, \eqref{expr2} and identity \eqref{confward} we find
\be
\partial_{\overline{z}} \,G(z,y) =0\,,
\ee
where 
$
G(z,y)\equiv \,\langle \Phi_K^\dagger(\infty) \phi_i(z) \,\,\phi_i^\dagger(y)\,\,\Phi_K(0)\rangle -{\overline{z}\over 
\overline{y}}\,\langle \Phi_K^\dagger(\infty) \chi_i^\dagger(z) \,\,\chi_i(y)\,\,\Phi_K(0)\rangle\,.
$
Setting $y=1$ we get \eqref{deffung} and \eqref{deffung2}. More details about analogous identities can be found in \cite{Chen:2026vml}.

\section{Some useful conformal block identities}

Consider the conformal block expansion of the correlator $\langle \Phi_K^\dagger(\infty)\,\, \phi_i(z) \,\,\phi_i^\dagger(1)\,\,\Phi_K(0)\rangle$ in the $z\rightarrow 0$ channel. The right-moving conformal blocks that will appear are of the form
\be
{\cal G}^{\widetilde{h}}(\overline{z}) = \overline{z}^{\widetilde{h}}\,\,_2F_1\left(\widetilde{h}-\widetilde{h}_K +{1\over 2},\widetilde{h}-\widetilde{h}_K +{1\over 2},2\widetilde{h},z\right)\,,
\ee
where $\widetilde{h}$ is the right conformal dimension of the exchanged operator. We notice that they obey the identity
\be
\label{cblockidentity}
{\cal G}^{\widetilde{h}}(\overline{z})  = \overline{z}\left[{\cal G}^{\widetilde{h}-1}(\overline{z})+{1\over 8}\left({(2\widetilde{h}_K-1)^2\over \widetilde{h}(\widetilde{h}-1)}+4\right){\cal G}^{\widetilde{h}}(\overline{z})+{((2\widetilde{h}_K-1)^2-4\widetilde{h}^2)^2\over 64 \widetilde{h}^2(4\widetilde{h}^2-1)}{\cal G}^{\widetilde{h}+1}(\overline{z})\right]\,.
\ee
Using this identity, we can recombine the two terms of \eqref{deffung} into a conformal block expansion, though the coefficients do not have to be positive. There are cancellations between long multiplets and the surviving terms give \eqref{g14bps}.

\section{Late time plateau at the orbifold point}

At the orbifold point, the torus two-point function of a single-trace untwisted operator ${\cal O}=\frac{1}{\sqrt{N}}\sum_{j=1}^N {\cal O}_{\text{seed}}^{(j)}$ is given by \cite{Belin:2025nqd}
\eqsp{
\left\langle {\cal O}(z_1) {\cal O}^\dagger(z_2)\right\rangle_\tau=\frac{1}{N N!}\sum_{\substack{g,h\in S_N\\gh=hg}}\Bigg[\sum_{\xi\in O(g,h)}G(\xi,\tau, {\cal O}_{\text{seed}}(z_1), {\cal O}_{\text{seed}}^\dagger(z_2))\prod_{\substack{\xi'\in O(g,h)\\ \xi'\neq\xi}}Z_{\text{seed}}(\tau_{\xi'})\\
+\sum_{\substack{\xi_1,\xi_2\in O(g,h)\\ \xi_1\neq\xi_2}}G(\xi_1,\tau, {\cal O}_{\text{seed}}(z_1))G(\xi_2,\tau, {\cal O}_{\text{seed}}^\dagger(z_2))\prod_{\substack{\xi'\in O(g,h)\\ \xi_1\neq\xi'\neq\xi_2}}Z_{\text{seed}}(\tau_{\xi'})\Bigg]\,.\label{eq:2pt}
}
The sums over $g,h$ encode the sum over cover surfaces, and $\xi$ encodes a connected component of such a cover surface. More precisely, $\xi$ is an element $O(g,h)$, the set of orbits of the abelian group generated by $g$ and $h$ acting naturally on the set $\{1,\dots,N\}$. The function $G$ encodes the operator insertions on such a connected component of a cover surface as
\eqsp{\label{eq:G}
G(\xi,\tau, {\cal O}_{\text{seed}}(z_1),\cdots {\cal O}_{\text{seed}}(z_n))&=\sum_{j_1,\dots,j_n=0}^{d_\xi-1}\sum_{\ell_1,\dots,\ell_n=0}^{a_\xi-1}d_\xi^{-n \Delta_{ {\cal O}}}\\
&\hspace{-60pt}\left\langle {\cal O}_{\text{seed}}\left(\frac{z_1+ 2\pi j_1+ 2\pi \ell_1\tau}{d_{\xi}}\right)\cdots  {\cal O}_{\text{seed}}\left(\frac{z_n+ 2\pi j_n+2\pi \ell_n\tau}{d_{\xi}}\right)\right\rangle_{\tau_{\xi}}\,,
}
Here $a_\xi$, $b_\xi$, and $d_\xi$ are integers that uniquely characterize the connected components of the cover surface, and the modular parameter $\tau_\xi$ is related to $\tau$ via
\eq{
\label{eq:tauxi}
\tau_\xi=\frac{a_\xi\tau+b_\xi}{d_\xi}\,.
}
Because our probe operator is untwisted, not all cover surfaces contribute; only primitive cover surfaces have the correct monodromy structure to contribute. This means that only terms with $a_\xi=1$ contribute in the sum over $\xi$. Moreover, $R$-charge conservation implies that only the first line of \eqref{eq:2pt} can be nonzero for a 1/2-BPS probe. Inserting a complete set of (seed) states on the cover and taking the late-time average of \eqref{eq:2pt} in the limit $\beta_R\rightarrow\infty$ results in \eqref{eq:orbplat}.

\end{document}